%% file: pks1136-135_mihai.tex
\documentclass[iop]{emulateapj}
\usepackage{url}
\usepackage{color}

\newcommand\alpharad{\alpha_{\mathrm{rad}}}

\newcommand\GHz{\;\mbox{GHz}}

\newcommand\kparsec{\ensuremath{\,\mathrm{kpc}}}

\newcommand\keV{\ensuremath{\,\mbox{keV}}}
\newcommand\SN{\ensuremath{\mathrm{S}/\mathrm{N}}}

\newcommand\etal{~et~al.}
\newcommand\lunderscore{\rule{1ex}{\fboxrule}}

\begin{document}

\title{Polarimetry and the High-Energy Emission Mechanisms in Quasar Jets: The case of PKS~1136$-$135.}
\shorttitle{Polarimetry of PKS~1136$-$135}

\author{Mihai Cara\altaffilmark{1,2}, Eric S. Perlman\altaffilmark{1}, Yasunobu Uchiyama\altaffilmark{3}, Chi C. Cheung\altaffilmark{4}, Paolo S. Coppi\altaffilmark{5}, Markos Georganopoulos\altaffilmark{6}, Diana M. Worrall\altaffilmark{7}, Mark Birkinshaw\altaffilmark{7}, William B. Sparks \altaffilmark{8},  Herman L. Marshall \altaffilmark{9}, Lukasz Stawarz\altaffilmark{10,11}, Mitchell C. Begelman\altaffilmark{12}, Christopher P. O'Dea\altaffilmark{13}, Stefi A. Baum\altaffilmark{14}}

\altaffiltext{1}{ Department of Physics and Space Sciences, Florida Institute of Technology, 150 W. University Blvd., Melbourne, FL 32901, USA}
\altaffiltext{2}{Current Address: Physics Department, Case Western Reserve University, 2076 Adelbert Rd., Cleveland, OH, 44106-7079}
\altaffiltext{3}{SLAC/KIPAC, Stanford University, 2575 Sand Hill Road, M/S 209, Menlo Park, CA 94025, USA}
\altaffiltext{4}{Space Science Division, Naval Research Laboratory, Washington, DC  20375, USA}
\altaffiltext{5}{Yale University, Department of Astronomy, PO Box 208101, New Haven, CT 06520-8101}
\altaffiltext{6}{Department of Physics, University of Maryland -- Baltimore County, 1000 Hilltop Circle, Baltimore, MD  21250, USA}
\altaffiltext{7}{Department of Physics, University of Bristol, Bristol, BS8 1TL, UK}
\altaffiltext{8}{Space Telescope Science Institute, 3700 San Martin Drive, Blatimore, MD 21218, USA}
\altaffiltext{9}{Kavli Institute for Astrophysics and Space Research, Massachusetts Institute of Technology, Cambridge, MA 02139, USA}
\altaffiltext{10}{Institute of Space Astronautical Science, JAXA, 3-1-1 Yoshinodai, Chuo-Ku, Sagamihara, Kanagawa 252-5210, Japan}
\altaffiltext{11}{Astronomical Observatory, Jagiellonian University, 30-244 Krakow, Poland}
\altaffiltext{12}{Department of Astrophysical and Planetary Sciences, UCB 391, University of Colorado, Boulder, CO  80309-0391}
\altaffiltext{13}{Laboratory for Multiwavelength Astrophysics, School of Physics and Astronomy, Rochester Institute of Technology, 84 Lomb Memorial Dr., Rochester, NY  14623-5603}
\altaffiltext{14}{Chester F. Carlson Center for Imaging Science, Rochester Institute of Technology, 54 Lomb Memorial Dr., Rochester, NY 14623-5604}

\begin{abstract}

Since the discovery of kiloparsec-scale X-ray emission from quasar jets, the physical processes 
responsible for their high-energy emission have been poorly defined.  A number of mechanisms are 
under active debate, including synchrotron radiation, inverse-Comptonized CMB (IC/CMB) emission, 
and other Comptonization processes.  In a number of cases, the optical 
and X-ray emission of jet regions are inked by a single spectral component, and in those, high-
resolution multi-band imaging and polarimetry can be combined to yield a powerful diagnostic of jet 
emission processes.  Here we report on deep imaging photometry of the jet of PKS~1136$-$135 
obtained with the {\it Hubble Space Telescope.}  We find that several knots are highly polarized in the 
optical, with fractional polarization $\Pi>30\%$.  When combined with the broadband spectral shape 
observed in these regions, this is very difficult to explain via IC/CMB models, unless the scattering 
particles are at the lowest-energy tip of the electron energy distribution, with Lorentz factor $\gamma 
\sim 1$, and the jet is also very highly beamed ($\delta \geq 20$) and viewed within a few degrees of 
the line of sight.  We discuss both the IC/CMB and synchrotron interpretation of the X-ray emission 
in the light of this new evidence, presenting new models of the spectral energy distribution and also the 
matter content of this jet.  The high polarizations do not completely rule out the possibility of IC/CMB  
optical-to-X-ray emission in this jet, but they do strongly disfavor the model.  We discuss the 
implications of this finding, and also the prospects for future work.


\end{abstract}


\maketitle

\section{Introduction}

The jets of radio-loud Active Galactic Nuclei (AGN) carry energy and matter out
from the nucleus to cluster-sized lobes, over distances of hundreds of kpc. 
While found in only $\sim 10\%$ of AGN, jets can have a power output (including
both luminosity and kinetic energy flux) comparable to that of their host galaxies
and AGN \citep{Rawlings:91}, and can profoundly influence the evolution of the
hosts and neighbors  \citep{MacNul:07}. AGN jets are completely ionized flows, and the
radiation we see from them is non-thermal in nature. The highest-power radio jets
typically terminate in bright ``hot-spots'' and are classified as Fanaroff-Riley
type~II (FR~II)   while the lower-power radio-jets are 
brightest at the core and are classified as Fanaroff-Riley type~I (FR~I)
 \citep{FR:74}. That the radio emission arises from synchrotron
radiation is supported by strong linear polarization and power-law spectra seen
in both lower- and higher-power jets. However, in the optical and X-ray bands,
the nature of the emission from higher-power large-scale jets
is less clear.

In low-power FR~I radio galaxies, the optical and X-ray fluxes fit on
extrapolations of the radio spectra \citep[see
e.g.,][]{Perlman:01,Hardcastle:01,Perlman:05}, and high polarizations are seen
in the optical \citep[typically $\sim 20-30\%$,][]{Perlman:99,Perlman:06}
suggesting synchrotron emission.  These jets exhibit a wide variety of
polarization properties \citep{Perlman:06,Dulwich:07,Perlman:10}, often
correlated with X-ray emission. For example, in the jet of M87
\citep{Perlman:05}, a strong anti-correlation between  the intensity of X-ray emission and optical
polarization fraction was found in the knots, accompanied by changes in the magnetic
field direction, suggesting a strong link between the jet's dynamical structure
and high-energy processes in the jet interior, where shocks compress the
magnetic field and accelerate particles. 

For the more powerful FR~II and quasar jets, the nature of both the optical and
X-ray emission is under active debate. 
The first issue to resolve is whether the mechanism and electron population that
produces optical emission is linked with the radio or X-ray emission.  For some knots in FR II jets,
the radio-to-X-ray SED is consistent with the optical and X-ray arising as synchrotron 
from the same electron population as that producing the radio emission \citep[see e.g.,][]{WorBir:05,Kraft05,Kataoka:08}.  In others, however, the optical emission
can lie well below an interpolation between the radio and X-rays
\citep[e.g.,][]{Sambruna:04}, sometimes by decades \citep[e.g.,
PKS~0637$-$752,][]{Schwartz:00,Mehta:09}, resulting in a characteristic
double-humped shape of spectral energy distribution. In some jets the optical
emission appears linked to the X-ray emission by a common component, as seen 
in both 3C~273 and PKS~1136$-$135, where deep Hubble Space Telescope (HST),
Chandra X-ray Observatory and infrared Spitzer Space Telescope imaging
\citep{Jester:01,Jester:06,Jester:07,Uchiyama:06,Uchiyama:07} has shown that a
second component, distinct from the lower-energy synchrotron emission, arises in
the near-IR/optical and dominates the jet emission at optical and higher
energies, at least up to $10\keV$. Competing mechanisms have been proposed: either
synchrotron radiation from very high-energy particles or inverse-Comptonization
of low energy photons off relativistic electrons of the jet
\citep[see][]{Kataoka:05,Harris:06,Worrall:09}, however, the nature of this component
cannot be constrained by multi-waveband spectra alone \citep{Georg:06}.


Polarimetry is a powerful diagnostic for jets because synchrotron emission is
naturally polarized, with the inferred direction of the magnetic field vector
indicating the weighted direction of the magnetic field in the radiating volume
and the fractional polarization indicating relative ordering of the magnetic
field. In FR~IIs, where the radio-optical spectrum often cannot be extrapolated
to the X-rays, high-energy synchrotron emission is a possibility, but often requires a second electron
population. For FR~IIs to accelerate particles to $\gamma > 10^7$, as
required for X-ray synchrotron emission,  requires highly efficient particle
acceleration mechanisms that can operate well outside the host galaxy (e.g., in
PKS~1136$-$135 the X-ray emitting knots are at projected distances of
$30-60\kparsec$ from the AGN).  If the optical and X-ray emission is
synchrotron radiation, the optical polarization will be high, comparable to that
seen in the radio, but with characteristics that may be linked to acceleration
processes.

The second possibility is inverse-Comptonization of Cosmic Microwave Background
photons \citep[IC/CMB,][]{Celotti:01}.  This requires a jet that remains highly
relativistic out to distances of hundreds of kpc
\citep{Schwartz:00,Tavecchio:00,Georg:03}, viewed at a small angle to the
line of sight.  Any optical IC-CMB would be linked to nearly cold electrons,
with $\gamma<10$, a population of particles that has never before been tracked.
If the emitter is moving at
relativistic bulk speeds, $\Gamma \gg 1$, then the forward-bunching effect will
make the CMB photons essentially unidirectional in the jet frame. The IC
scattering on the unidirectional and unpolarized photon beam by high-energy
electrons having large Lorentz factors ($\gamma \gg 1$) should be unpolarized.
On the other hand, the scattered radiation from cold electrons ($\gamma \sim 1$;
so-called vulk Comptonization) in the jet is expected to be highly polarized
\citep{Begelman:87}.  \citet{Uchiyama:11} have carried out calculations covering
the intermediate regime with $\gamma \sim$~few, making use of the general
expression for the intensity and polarization of singly-scattered Comptonized
radiation presented by \cite{Poutanen:93}. They found that, for a power-law
energy distribution of electrons with a cutoff at $\gamma_{\min} = 2$, the
polarization degree can be as large as $8\%$ with the direction of the electric
field vector perpendicular to the jet axis \citep[see also][]{McNamara:09}.

It is also possible to Comptonize other photon fields.  The most commonly 
cited process is synchrotron
self-Compton (SSC) radiation, in which the seed photons come from the jet's
low-frequency radio emission.  While SSC is unavoidable, it is unlikely to
dominate the X-ray emission of the jet knots because in order to fit the
observed X-ray emission, one requires a jet that is massively out of
equipartition (by factors of $20-100$), has large viewing angle and/or is
unbeamed.  However, SSC is the leading scenario for X-ray emission from the
terminal hotspots of the most powerful jets \citep{Harris:94,Wilson:01}. SSC predicts
optical polarization properties similar to that of the lowest-frequency radio
emission.

All of these issues have important implications on the overall mass-energy budget
and energetics of powerful jets.  For example, a large population of nearly cold electrons
 could carry the vast majority of the matter content
and energy budget of the jet, making them vital in constraining the overall
energetics of the jet.  A second factor, which is even more difficult
to constrain, would be whether the jet flow is composed purely of leptons, or includes hadrons \citep[e.g.,][]{Georg:05}. If either of these components is found, the jet flow could become more powerful by orders of magnitude, resulting in a much greater impact on the surrounding ICM and more extreme demands on the overall physics
\citep[e.g.,][]{Ghis:01}.

Up until now the only existing optical polarimetric observations of quasar jets
were those of the quasar 3C~273, where ground based polarimetry
\citep{Roser:91} yielded significantly different results from
(pre-COSTAR FOC) space-based observations with HST \citep{Thomson:93}.  The discrepancy
could not be reconciled by simply accounting for differences in resolution. In
this paper we present, for the first time, high-resolution space-based optical
polarimetric observations of the quasar jet PKS~1136$-$135.

PKS~1136$-$135 is a steep-spectrum quasar at $z=0.556$.  It has a low integrated 
optical polarization $<1\%$ \citep{Sluse:05}.  
It is part of the 2 Jy sample \citep{WP:95}, and spectroscopy and images of its host galaxy
were presented by \citet{Tadhunter:93} and \citet{Ramos:11} respectively. 
It was selected for inclusion in a search for X-ray jet sources based on the brightness of 
its radio jet by \citet{Sambruna:02}, whose short X-ray observation with {\it Chandra} 
was supplemented with a wide-band {\it HST} image.  Later, deeper {\it Chandra} and {\it HST}
observations were analyzed by \citet{Sambruna:06}.  Its X-ray jet features emission from 
five regions of the jet, plus the terminal hotspot.  
Modeling of the jet SED by \citet{Sambruna:06}, which assumed IC/CMB for the jet X-ray emission mechanism,  found a Doppler factor $\delta \sim 7$.  Thanks to data from the {\it VLA, HST}
and {\it Chandra}, as well as {\it Spitzer}, the spectral energy distribution (SED) of its jet is the best sampled
among the lobe-dominated quasars.  The jet has a morphology similar to that seen in 3C~273, displaying an
anti correlation between radio and X-ray flux.  This has been interpreted as being indicative of
deceleration \citep{Sambruna:06,Tavecchio:06}. 

The paper is laid out as follows.
In Section~2, we describe the observations and data reduction methods. 
Section~3 describes the morphology, polarimetry and spectral imaging results we obtain
for PKS~1136$-$135.  Section~4 describes model fitting and a discussion of the implications
of these results. Finally we close in Section~5 by stating our conclusions.
Throughout this paper we assume a cosmology with $\Omega_m=0.27$,
$\Omega_{\Lambda}=0.73$, $\Omega_{r}=0$ and $H_0=71\mbox{ km}\mbox{
s}^{-1}\mbox{ Mpc}^{-1}$.  For PKS~1136$-$135 this leads to a luminosity distance of 3.20~Gpc, 
and an angular scale of $1''$ = 6.4 kpc.
We also adopt the following convention for the
spectral index, $\alpha$: $F_{\nu}\propto\nu^{-\alpha}$, where $\nu$ is the
frequency.

\section{Observations and Data Reduction}

\subsection{Polarimetric Observations}

Optical polarimetry of the jet of the quasar PKS~1136$-$135 was performed
(proposal \#11138, Cycle~16) with HST's Wide Field and Planetary Camera~2
(WFPC2) between March 2 and March 16, 2008 using the Wide Field (WF) chips of the
WFPC2 camera with the F555W (broadband $V$) and POLQ filters. To reconstruct the
Stokes parameters, it is necessary to observe a source using different polarizer
orientations. However, as the polarizer can rotate through only 51$^\circ$, rotating 
the polarizer results in a very small field of view.  We therefore followed the practice of 
\citet{Perlman:99,Perlman:06} and obtained images in all three WF chips (WF2, WF3,
and WF4) for which the polarizer has a nominal orientation of $0\arcdeg$,
$45\arcdeg$, $90\arcdeg$ respectively, where an orientation of $0\arcdeg$ lies
roughly along the $+X$ direction of the PC1 chip (This corresponds to values of $45\arcdeg$,
$90\arcdeg$, and $135\arcdeg-PA\lunderscore V3$ with respect to North, where
$PA\lunderscore V3$ is the angle between North and the $V3$ axis of the
telescope; see Biretta \& McMaster 1997 for more details). The observations were
performed in 21 orbits (exposure time of $2500$ seconds per orbit) evenly split
between WF chips for a total of $17500$ seconds of exposure time per polarizer
orientation giving us the deepest optical image of the PKS~1136$-$135 jet to
date. We used a simple line-dither pattern for our
observations which allowed for a better recovery of information in the
defective/hot pixels and pixels affected by cosmic rays (CR). The F555W filter
was chosen for the following reasons: i) based on pre-existing photometry, 
its pivot wavelength falls inside the ``dip'' between the low-energy
(radio to infrared) and the high-energy (optical to X-ray) bumps of the
broadband spectrum thus minimizing the contribution from the low-energy
(synchrotron) component to the total flux and polarization; ii) it complements
existing non-polarimetric observations at other optical wavelengths \citep[{\it HST} F475W, F625W, F814W;
see][]{Sambruna:06}; iii) it provides optimal performance of the POLQ
filters (i.e., maximized parallel and minimized perpendicular transmissions).

\subsection{Image Processing}

As the first step, we re-calibrated the images using the most up-to-date
reference files (i.e., flat field files, distortion correction table, etc.)
obtained from the STScI Calibration Database System. In addition to this
standard calibration procedure, we have re-computed the 
PHOTFLAM and PHOTPLAM keyword values using the \verb|calcphot| routine from the \verb|SYNPHOT|
package assuming a power-law spectral distribution with a spectral index
$\alpharad=1$. We used the \verb|Multidrizzle| task \citep{Fruchter:09} from the
\verb|STSCI_PYTHON| package to drizzle-combine the images for each polarizer
orientation (WF chip). Besides combining the images, \verb|Multidrizzle|
distortion-corrects the images, performs image flat-fielding, cosmic-ray
rejection, image alignment, and other tasks.

However, in order to obtain good final drizzle-combined images, it is necessary
to supply good image alignment information to the \verb|Multidrizzle| task. To
find the necessary geometric transformations we used an iterative process so
that after each iteration we have obtained an improved estimate for image shifts
and rotations. This process was done as follows. At first, we individually
drizzled all images onto a common frame and, using the positions of the same
star-like object in each drizzled image, we found an {\it initial} estimate of
the shifts necessary to align the images (no rotations or distortions were considered
at the initial stage).  We then set up the following iterative process:
\begin{enumerate}
	\item[i)] we use \verb|Multidrizzle| with the available geometric
transformations (in the form of a ``shifts file'') to perform distortion
correction and clean cosmic rays from the images;
	\item[ii)] on these images we run the  \verb|Tweakshifts| task (a part
of the \verb|STSCI_PYTHON| package) to find the \emph{delta}-shifts and
rotations between images using 13 reference objects (stars and unresolved
clusters). \verb|Tweakshifts| automatically excludes the objects that cannot be
fit well with a best fit model for geometric transformations;
	\item[iii)] we update the old shifts file with delta-shifts found in the previous step and repeat this process starting with step i) until we obtain corrections to shifts and rotations smaller than $10^{-3}$~pixels and $10^{-4}$~degrees accordingly.
\end{enumerate}
Decreasing values of the corrections are an indication of convergence of the iterative process. However, these corrections are not an indication of goodness of alignment.  For this, we use the root mean square of the residuals of the fit as reported in the transformations database created by the \verb|geomap| task which is used internally by the \verb|Tweakshifts| task. More precisely, we define misalignment error as
\begin{equation}
  \sigma_\mathrm{fit}=\sqrt{\frac{1}{N-1}\sum_{i=2}^N((\sigma^i_{x,\mathrm{fit}})^2+(\sigma^i_{y,\mathrm{fit}})^2)},
\end{equation}
where $\sigma^i_{x,\mathrm{fit}}$ and $\sigma^i_{y,\mathrm{fit}}$ are coordinate
residuals of the fit as reported in the \verb|geomap|'s database file. Index $i$
numbers input images to the \verb|Tweakshifts| task. Since the alignment is
performed relative to the first input image, the summation in the above equation is
performed over index $i$ that runs from $2$ to $N$. For WFPC2 F555W$+$POLQ data
the estimated misalignment error was $0.16$~pixels and at this level it was shown
\citep{Perlman:06} to have a minor effect on polarimetry (this effect is further
minimized in the case of \emph{aperture} polarimetry). By performing the image
alignment procedure \emph{simultaneously} on all input images regardless of
their epoch or polarization filter, we avoid an extra step of
aligning drizzled images from the three polarizers (images with different
polarizer filters must be aligned to obtain Stokes images later) -- a step that
would have led to degradation of the image quality.

With the shift information from the image alignment step, we then combined
images with the same POLQ filter using the \verb|Multidrizzle| task. At first,
we performed sky background subtraction on the input images using our own
routine that can use selected regions of the sky and cosmic ray masks for
background evaluation. We then used the \verb|Multidrizzle| task to perform
distortion correction of the images, cosmic ray cleaning, rotate images
North-up, and drizzle-combine images into a single final image.

\subsection{Polarization Images}\label{sec:polimg}

Observations with different polarizer orientations (i.e., different apertures
with the POLQ filter) can be linearly combined \citep[see][]{Biretta:97} to
produce Stokes~$I$, $Q$, and $U$ images:
\begin{equation}\label{eq:polmatrix}
  \mathbf{S}=M\mathbf{W},
\end{equation}
where $\mathbf{S}=(I,Q,U)$ is a vector of Stokes images and
$\mathbf{W}=(W_2,W_3,W_4)$ is a vector of WFPC2 images corresponding to
different polarizer orientations (in our case WF2, WF3, and WF4 CCD chips with
POLQ filter). Coefficients of the Mueller matrix $M$ were computed using the
WFPC2 Polarization Calibration
Tool\footnote{\url{http://www.stsci.edu/hst/wfpc2/software/wfpc2\_pol\_calib.html}}.
Prior to combining the images using matrix $M$, it is necessary to check that
all images have similar resolution (differences in resolution could be due to
telescope optics, quality of shifts used by \verb|Multidrizzle|, jitter, and
other factors). This was done by measuring the Gaussian FWHM of 3--4 sharpest
globular clusters in the drizzle-combined images using the \verb|imexamine| task
and taking their average value. The WF4 image had the largest averaged FWHM
equal to $0\arcsec .1889$. Therefore, we applied a Gaussian blur filter of
appropriate standard deviation to the WF2 and WF3 images so as to make their average
FWHM match that of the WF4 image. Finally, we combine the $\mathbf{W}$ images
to obtain Stokes images.  The resulting image is shown in Figure~1.

As the quasar and host galaxy in PKS~1136$-$135 are quite bright, 
their emission almost completely hides the innermost jet knot. 
Therefore, it is important to perform galaxy subtraction prior to any
polarimetric measurements, and also mask out the diffraction spikes. 
Galaxy subtraction was done on the Stokes~$I$ image. First, we
masked out the jet, clusters, and other sources of emission except for the host
galaxy. We then used the \verb|ellipse| task to fit elliptical isophotes to the
host galaxy image and finally, we used the \verb|bmodel| task to create a
Stokes~$I$ image of the host galaxy from the fitted isophotes. To produce images
of the host galaxy that can be subtracted from the drizzle-combined images, we
set components $Q$ and $U$ of the vector $\mathbf{S}$ to $0$ (host galaxy
emission is unpolarized) and invert equation~(\ref{eq:polmatrix}) which will
produce three images of the host galaxy as if they were observed in CCD chips
WF2, WF3, and WF4. We then subtract these images from the corresponding
drizzle-combined image using the \verb|imcalc| task. Finally, we repeat polarization
combination (eq.~\ref{eq:polmatrix}), this time with the galaxy-subtracted
images as inputs. We combine the resulted Stokes~$I$, $Q$, and $U$ images in a
standard way to produce polarization ($P$) defined as $P=\sqrt{Q^2+U^2}$,
fractional polarization ($\Pi$) defined as $\mathrm{\Pi}=P/I$ and electric vector
position angle ($\Xi$) defined as $\mathrm{\Xi}=\frac{1}{2}\tan^{-1}(U/Q)$ (or,
alternatively, magnetic field position angle (MFPA) defined as
$\mathrm{MFPA}=\mathrm{\Xi}+90\arcdeg$).  

The errors in the polarization images have been computed using standard error
propagation with the errors of the input galaxy- and background-subtracted
images estimated as the sum, in quadrature, of the Poisson errors in the
original (prior to galaxy and background subtraction) images, standard deviation
of the background level (post galaxy subtraction), readout noise, and
quantization noise. For this analysis we have ignored any errors in the galaxy
model. We also added in quadrature $3\%$ absolute error to fractional
polarization to account for uncertainties in polarization calibration
\citep{Biretta:97}. We accounted for the well-known Rician bias in $P$
\citep{Serkowski:62} using a Python code adapted from the STECF IRAF package
\citep{Hook:00}. This code debiases the $P$ image following \citet{Wardle:74},
and calculates the error in PAÄ accounting for the non-Gaussian nature of its
distribution \citep[see][]{Naghizadeh:93}. In performing this calculation,
pixels with signal to noise (S$/$N)~$< 0.1$ were excluded outright, and since
the debiasing is done with a "most-probable value" estimator, pixels where the
most probable value of $P$ was negative, or above the Stokes~$I$ value (i.e., $\Pi
> 100\%$) were blanked. This code was first used in \citet{Perlman:06}.

\subsection{Aperture Polarimetry and Photometry}\label{sec:AperturePol}

\begin{figure}
 \includegraphics[height=.36\textheight,angle=-90]{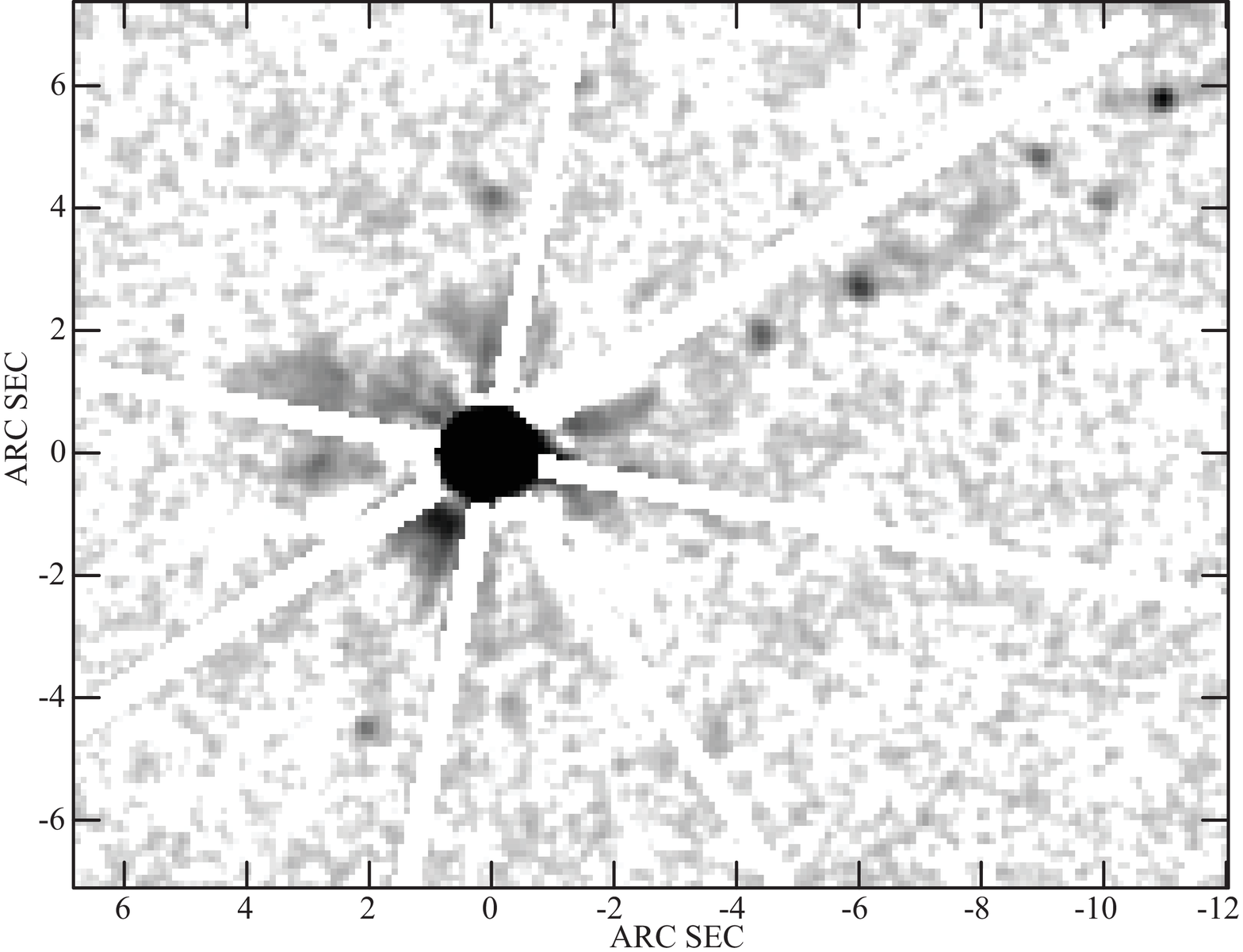}
  \includegraphics[height=.36\textheight,angle=-90]{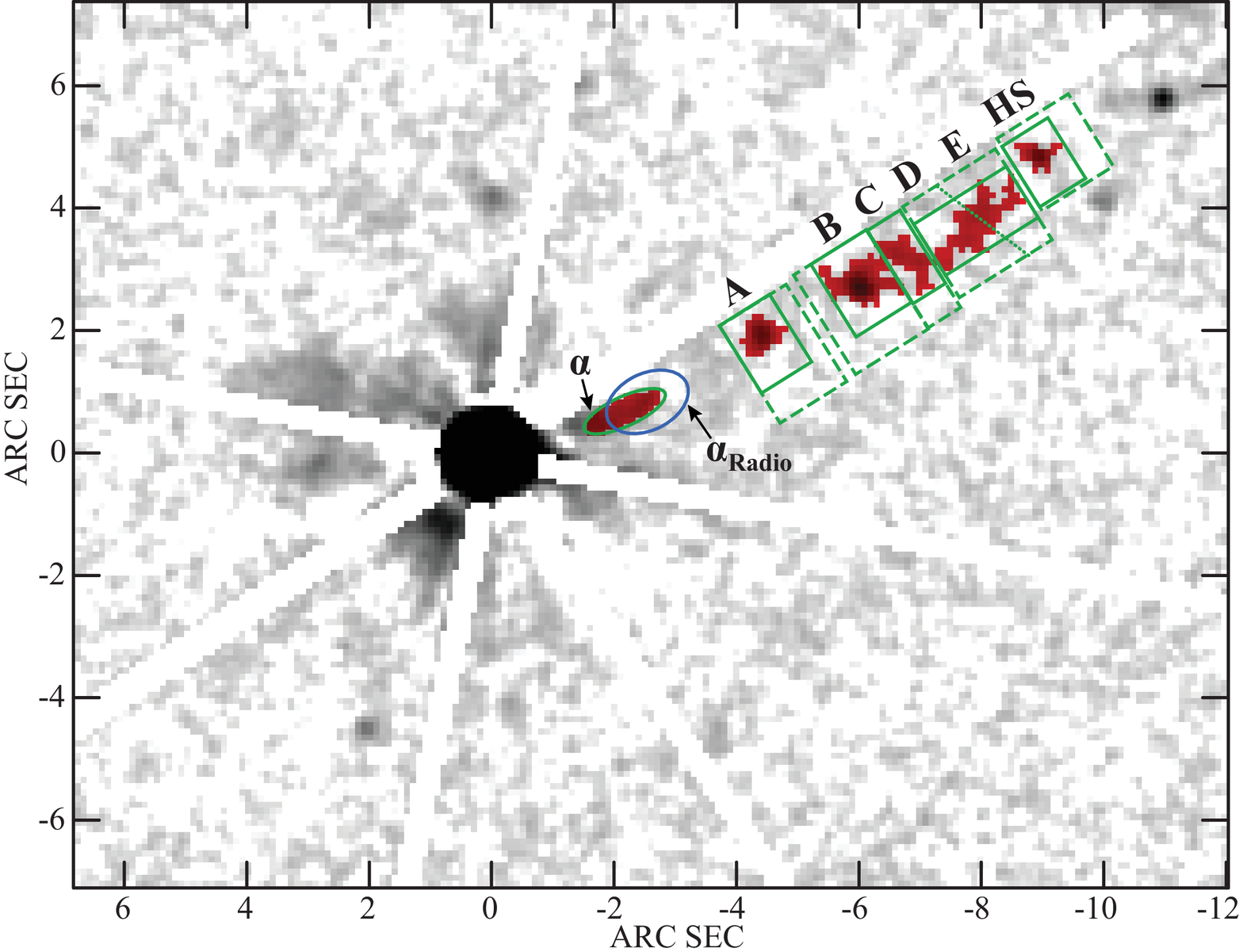}
  \caption{Galaxy subtracted HST F555W Stokes~$I$ image of the PKS~1136$-$135 jet, at top,
  and at bottom,  with
knot nomenclature and used apertures superposed. The white regions represent diffraction 
spike features that have been excluded from our analysis. Dashed green rectangles show the largest
apertures. Solid green rectangles and ellipse show smaller apertures.   These represent the regions
within which pixels were considered for inclusion in various knot regions. The
dotted green diagonal line separates knots~D and~E based on a radio image. The
$1\sigma$-cut apertures are shown as red masks overlaid on the jet knots. The
blue ellipse shows the location of the knot~$\alpha$ in the radio image. See \S\S~2.4 and~3.1 for discussion.}
\label{fig:apertures}
\end{figure}

Aperture polarimetry was performed using rectangular and elliptical regions
large enough to include most of the emission from the optical knots. In
Figure~\ref{fig:apertures} we show galaxy-subtracted Stokes~$I$ image of the
quasar jet PKS~1136$-$135 with green rectangles and a ellipse showing the
apertures used in this analysis together with knot nomenclature. Positions and
sizes of the apertures have been chosen so that they include the knots but avoid
the nearby saturated column, diffraction spikes, and other sources of signal.
The larger apertures include most of the flux and thus require minimal aperture
correction making the flux values more reliable.
However, large apertures tend to produce larger Poisson errors 
due to inclusion of many low $\SN$ pixels. On the other
hand, smaller apertures, while producing higher $\SN$ results, are more
susceptible to errors in aperture correction. In Figure~\ref{fig:apertures} the
dashed apertures show the largest apertures used while the continuous line
apertures show the location of smaller apertures around the knots. In order to
increase the $\SN$ of aperture measurements, we have also used even smaller
apertures produced by selecting only 4-point connected pixels with
$\SN>1\sigma$,  $\SN>2\sigma$, and  $\SN>3\sigma$ (we will call these apertures
$1\sigma$-cut, $2\sigma$-cut, and $3\sigma$-cut apertures accordingly). Inner
holes (if any) are filled-in so that the apertures are simply connected. In
Figure~\ref{fig:apertures} we show the 4-point connected apertures with
$\SN>1\sigma$ as red patches. Whenever possible, we have used several apertures
of varying size for each knot. For a given knot, the final value for the flux is
computed as a weighted mean \citep{Gough:08}: 
\begin{equation}
I=\sum_k w_k I_k
\end{equation}
and with an error estimate given by:
\begin{equation}
\label{eq:weightedERR}
\sigma_I=\sqrt{\frac{1}{1-\sum_k w_k^2}\sum_k w_k (I-I_k)^2},
\end{equation}
where summation is performed over apertures of different sizes. The weights were chosen as
\begin{equation}
\label{eq:weights}
w_k=\frac{1}{(\sigma_k C_k)^2} \left(\sum_k \frac{1}{(\sigma_k C_k)^2} \right)^{-1},
\end{equation}
such that more weight is given to the apertures with smaller measurement (e.g., Poisson errors, read-out noise, etc.) errors 
($\sigma_k$) and less weight to the apertures with large aperture and CTE correction factors ($C_k$) described below. The 
weights in equation~(\ref{eq:weights}) have been normalized such that $\sum_k w_k = 1$.  Extensive testing showed that 
the error estimates produced using this method were well justified.

Aperture correction was performed in the following way. We model point-like
knots (A, B, and the Hot-Spot) using a PSF generated with the TINYTIM PSF simulation
tool\footnote{\url{http://www.stsci.edu/software/tinytim/}}
\citep[see][]{Krist:93,Krist:04}. First, we generated an oversampled ($\times 4$)
PSF by providing TINYTIM with knot position, WFPC2 chip, used filter, jitter
information extracted from jitter file headers (\verb|jit| files), and set the
source spectrum to be a power law with spectral index $\alpha=1$. Then we rotated and
shifted the PSF to the same position and orientation as the knot of interest,
re-sampled the PSF to the Stokes~$I$ image sampling, and convolved it with a
Gaussian kernel with a standard deviation chosen such that the blurred PSF has
the same FWHM as the knot (measured with the \verb|imexamine| task). After
normalizing the PSF so that total flux is equal to unity, the aperture
correction is simply equal to the flux in the PSF that is outside the aperture
of interest. With this method we are able to apply aperture corrections to all
apertures. For extended knots ($\alpha$, C, D, and E), which cannot be modeled
with the PSF, we use $2\sigma$-cut apertures as model of the knots assuming a
constant value for the model flux. We then convolve these models with PSF with a
FWHM equal to the FWHM of the unresolved clusters in our image ($0\arcsec
.1889$, see \S~\ref{sec:polimg}). For these extended knots, the smallest
apertures for which we perform aperture correction are the $1\sigma$-cut
apertures. Because of the proximity of the knots~$\alpha$ and~$\alpha_{\mathrm{Radio}}$ to the core and diffraction spikes, we use only the $1\sigma$-cut apertures for flux measurements in these knots.

We correct for charge transfer efficiency (CTE) 
losses using \citet{Dolphin:09} formulae, which are valid for
point sources. We then estimate the CTE correction for extended sources
following the recommendations from ``WFPC2 Phase II Observation
Strategies''\footnote{\url{http://www.stsci.edu/hst/wfpc2/wfpc2\_phase2.html}},
more specifically, we divide CTE loss values obtained using \citet{Dolphin:09}
formulas by the extent (in native detector pixels) of the knots which we
estimate as $\frac{1}{2}\sqrt{N_{\mathrm{pix}}}$, where $N_{\mathrm{pix}}$ is
the number of pixels in an aperture.

Aperture and CTE corrections are applied to the count-rates in the apertures from each WF CCD 
image, which are then combined using the same prescription as in \S~\ref{sec:polimg} to compute 
Stokes parameters which, in turn, are used to compute $P$, $\Pi$, and EVPA. We correct for the 
Rician bias and compute errors on polarimetric quantities in a similar fashion to \S~\ref{sec:polimg}. 
For aperture polarimetry we use the smallest $\sigma$-cut apertures for which we could perform 
aperture correction ($3\sigma$-cut apertures for point-like knots~A, B, and the Hot Spot, and $1\sigma
$-cut apertures for the extended knots). For knots in which the smallest apertures gave zero 
polarization after debiasing, we used next non-zero aperture for that knot to estimate the upper limit of 
polarization at the $2\sigma$ level.

Normally, to convert a Stokes~$I$ count-rate image (or aperture measurements) to
flux units we would multiply the count-rate values by the value of the flux unit
conversion header keyword PHOTFLAM ($U_\lambda$). However, images from different
WF CCD chips have different PHOTFLAM and PHOTPLAM (pivot wavelength,
$\lambda_P$) values, which are also weakly dependent on spectral shape. Since
the WFPC2 Polarization Calibration Tool \citep{Biretta:97} assumes
count-rates as inputs, we cannot convert input images to flux units and then
combine them using equation~(\ref{eq:polmatrix}). To deal with this problem we
adopt the following strategy of computing ``average'' PHOTFLAM and PHOTPLAM
values. Let $M_{I,k}$ be the coefficients of the matrix $M$ (see
eq.~\ref{eq:polmatrix}) used to combine WF2, WF3, and WF4 images (index $k$ runs
from 2 to 4) into Stokes~$I$ image. Following the expressions for the flux unit
conversion factor and pivot wavelength from the ``Synphot User's Guide''
\citep[see \S~7.1 of][]{Laidler:05}, we define ``average'' PHOTFLAM and PHOTPLAM
values as:
\newlength{\oldarraycolsep}
\setlength\oldarraycolsep{\arraycolsep}
\setlength\arraycolsep{0.2em}
\begin{eqnarray}
  \left\langle \lambda_P \right\rangle  =  \sqrt{B/A},  &&  \left\langle U_\lambda \right\rangle  =  \frac{1}{B} \sum_{k=2}^4 M_{I,k},
\end{eqnarray}
where:
\begin{eqnarray}
  A = \sum_{k=2}^4 \frac{M_{I,k}}{\lambda_{P,k}^2 U_\lambda},  &&  B =  \sum_{k=2}^4 \frac{M_{I,k}}{U_\lambda}. \nonumber
\end{eqnarray}
We use these averaged PHOTFLAM and PHOTPLAM values to convert Stokes~$I$
count-rates to flux units. We also add in quadrature $2\%$ error to the total
flux error (see eq.~\ref{eq:weightedERR}) to account for uncertainties in
photometric calibration \citep{Baggett:02}.
\setlength\arraycolsep{\oldarraycolsep} 

\subsection{{\it HST} Photometry With F475W, F625W, and F814W}

{\it HST} observations and photometry of the PKS~1136$-$135 jet with F475W, F625W, and
F814W filters were performed by \citet{Sambruna:06} (proposal \#9682,
Cycle~11) using ACS/WFC. We have reprocessed these images and used the data to place better
constraints on the optical spectrum and broad-band SED of the jet components. However, due to the short
exposure time of these observations ($676\:\mathrm{s}$ for F475W and F625W
observations, and $520\:\mathrm{s}$ for F814W observations), the $\SN$ in the
extended knots is too low to reliably place apertures based on HST images.
Because of this, the apertures used for HST photometry in \citet{Sambruna:06}
were centered on X-ray/Radio positions. This may be a problem if the location of
X-ray/Radio knots is different from the location of the optical knots. In
Figure~\ref{fig:apertures} the blue ellipse shows the location of the innermost
knot~$\alpha_\mathrm{Radio}$ in the radio image (compare with the green ellipse
showing the knot~$\alpha$ in our deep F555W HST image) and it is clear that
optical knot~$\alpha$ and radio knot~$\alpha_\mathrm{Radio}$ have different
locations. Because of a much higher $\SN$ (exposure time $17500\:\mathrm{s}$ per
polarizer) of our {\it HST} observations, we decided to re-process the earlier  
F475W, F625W, and F814W observations using the procedure of \S~\ref{sec:AperturePol}) and 
apertures as defined in Figure~1.

Since processing of these images followed the same methodology as for our F555W data, we
mention here only the differences. First, because F475W, F625W, and F814W
observations were CR-SPLIT (and not dithered as were our F555W observations),
there was no need to find shifts between input images. Secondly, since the ACS
WFC detector's pixel scale is $0.05\arcsec$ compared to $0.0996\arcsec$ of the
WF detectors, we used the \verb|geomap| and \verb|geotran| tasks to re-scale
drizzle-combined images to match the scale of the F555W observations and 
to align these images to the F555W Stokes~$I$ image. Unfortunately,
because the F475W, F625W, and F814W observations had only two exposures (input
images) per filter, \verb|Multidrizzle| was unable to remove all the cosmic rays
from the input images.   
When these CR where located in some knots, we removed the corresponding
pixels from that knot's aperture mask. Also, because of the lack of dithering, the F475W, F625W and
F814W data have a significant number of warm/hot pixels and other image defects.
Finally, for CTE loss correction we used
ACS-specific formulas from the ``ACS Data Handbook'' \citep{Pavlovsky:05}.

\input{table1.tex}

\input{table2.tex}

\input{table3.tex}

\subsection{{\it Chandra} X-ray Observations}

Observations of the PKS~1136$-$135 jet were obtained with the {\it Chandra}
X-ray Observatory on 2003 April 16 (ObsID 3973) by R. Sambruna and collaborators
\citep{Sambruna:06}.
The total exposure was 77.4 ks.  The observations were obtained
with ACIS-S, with the sources at the aim point of the S3 chip.  
The 1/8 subarray mode was used, with a frame time of 0.4s, to minimize the 
effect of pileup from the quasar itself.  The source was also observed at a range of 
roll angles to place the jet away from the charge transfer tail of the quasar nucleus
and avoid flux contamination. 

We re-reduced the observations using CIAO version~4.2, with standard screening 
criteria and calibration files provided by the {\it Chandra} X-ray Center.  
Pixel randomization was removed, and only events in grades 0, $2-4$ and 6 were 
retained.  We also checked for flaring background events.   In order to aid 
comparison to the {\it HST} data, we subsampled the native {\it Chandra} resolution 
by 5, leading to a pixel scale of $0.0984''$/pixel.  In order to maximize the ability to 
detect low-level extended emission, we smoothed the observations adaptively using 
$csmooth$ in CIAO, requiring each cell to have a minimum of 10 photons.  

\subsection{Radio Observations}

We obtained NRAO\footnote{The National Radio Astronomy Observatory is a facility of the 
National Science Foundation operated under cooperative agreement by Associated
Universities, Inc.} Very Large Array (VLA) observations of PKS~1136$-$135 at 8.5 and $22\GHz$.  
At $22\GHz$, about 2.3~hrs 
total integration was obtained during a 24 hr observing run from 2002 May 27-28 in the hybrid 
BnA-array (program AC641). A full description of these data and the reduction procedures was 
presented in \citet{Cheung:04}. At $8.5\GHz$, we obtained about 7.5 hr of exposure on 
PKS~1136$-$135 on 2003 November 8 in the  B-array 
(AC689). The total intensity data from both frequencies were published in 
\citet{Sambruna:06} and \citet{Uchiyama:07}, 
and the polarization data are newly presented here to compare to the optical results. 
The phase calibrator used in both experiments was 1127$-$145 and the flux density 
scale was set using 3C~48 ($22\GHz$) and 3C~286 ($8.5\GHz$). For polarization calibration, 
leakage terms 
in the $22\GHz$ observing run were calibrated using observations of two bright point dominated 
sources (1354$+$195 and 3C~454.3) observe;d over a wide range of parallactic angles. In the $8.5\GHz$ run, the leakage terms were determined using the unpolarized source OQ208 and were 
found to be consistent with those derived from the 1127$-$145 scans. Both experiments used 
observations of 
3C~286 to set the absolute electric vector position angle (EVPA).

\subsection{{\it Spitzer} Infrared Observations}

Photometric data for the jet knots with the  \emph{Spitzer} IRAC~3.6 and~5.8 $\mu$m arrays 
are taken from \citet{Uchiyama:07}, in which IRAC observations carried out 
on 2005 June 10 (\emph{Spitzer} program ID 3586) were analyzed.
 Since the separation of adjacent 
knots is typically $\sim 1\arcsec$, comparable with the PSFs of the IRAC at~3.6 $\mu$m, the infrared 
images were fitted with a series of the PSFs located at the knots, after subtraction of 
the PSF wings of the quasar core. 
 It was difficult to 
measure fluxes individually from knots~C, D, and~E, so a combined flux was reported 
for them (referred to as knot~CDE) in \citet{Uchiyama:07}.

\section{Imaging Results}

\subsection{Jet Morphology and Aperture Photometry}

Our F555W Stokes I image is a factor 
of several deeper than the {\it HST} images previously obtained by \cite{Sambruna:06}.  
A total of seven knot regions are firmly detected in the optical.  We list flux densities for 
these regions in Table~1, not only in our F555W image but also in other {\it HST} observations (\S~2.5) 
as well as  in archival  observations with the {\it Chandra} X-ray Observatory (\S~2.6), VLA (\S~2.7) 
and  {\it Spitzer} Space Telescope (\S~2.8).  Each of the archival datasets was published previously,
with photometry \citep{Sambruna:06, Uchiyama:07}.  However,   
our results differ significantly from those published in \cite{Sambruna:06}.  This includes 
improved region definition and galaxy and PSF subtraction (which was not done by those authors).  
These have led us to improve significantly on the results of \citet{Sambruna:06} using their data.  
We can confidently claim detection of all knot regions in the F814W images,
whereas \citet{Sambruna:06} only published $3\sigma$ upper limits for some. 
We agree with \citet{Sambruna:06}'s result of non-detections of most of the fainter knots 
in F625W and F475W.   In Table~2, we give optical polarization properties
for the jet components in apertures shown in Figure~1.  Table~3 gives spectral indices
in radio, optical and X-ray for these jet features.

Figure~1 shows the {\it HST} image with the regions used for photometry overplotted.
In Figure~2a, we show the {\it HST} polarization image of PKS~1136$-$135 as greyscale.  The 
radio $8.5\GHz$ image is shown in contours, with polarization vectors representing the 
degree of polarization and direction of the inferred magnetic field.  The
contours and vectors are shown with the $8.5\GHz$ radio image overplotted in greyscale.
The polarization features in this image are discussed in the next sub-section.   Figure~2b shows the inverse of this overlay, with the radio image in greyscale and contours
and vectors representing the degree of polarization and inferred magnetic field direction
seen on the {\it HST} image.  Figure~3 shows the {\it Chandra} X-ray image in color, with contours from the radio (blue) and
{\it HST} (green) images overplotted.  Finally, Figure~4 shows an overlay of the {\it HST} F555W
image with the higher-resolution, $22\GHz$ radio image.  These overlays enable 
us to make the first high-quality assessment of the optical morphology of the knots and compare
them in detail to what is seen in other bands.

\begin{figure}
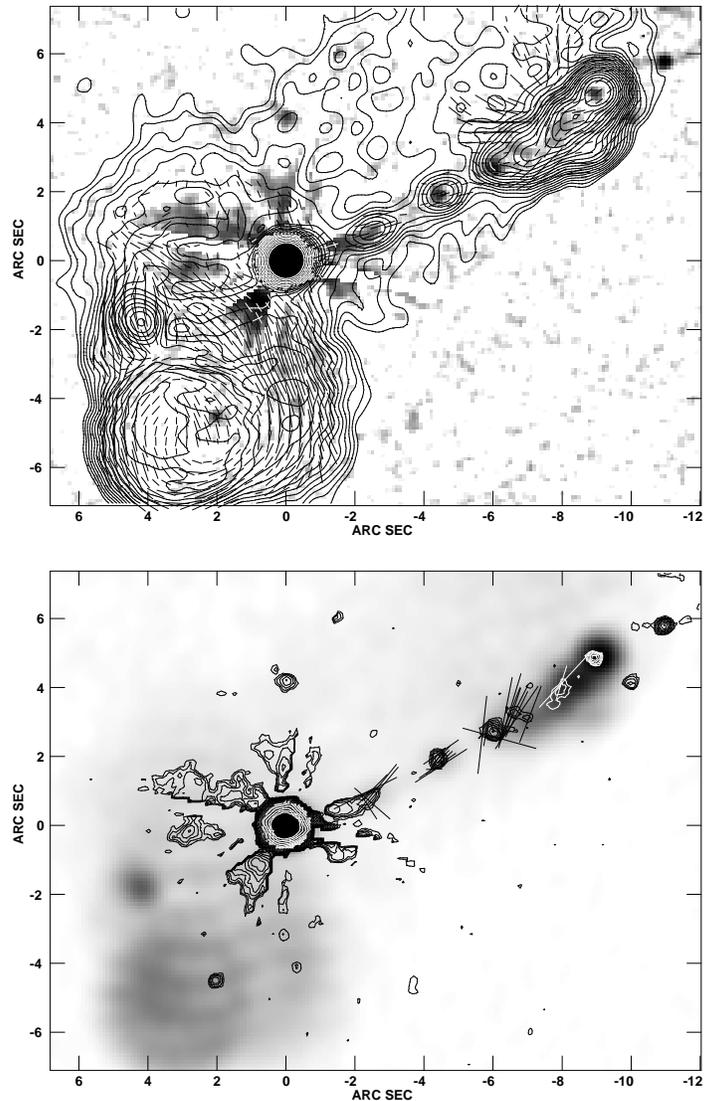

  \includegraphics[height=.4\textheight,angle=-90]{f2a.eps}

  \includegraphics[height=.4\textheight,angle=-90]{f2b.eps}
  \caption{At top (Figure~2a), the Galaxy subtracted {\it HST} F555W Stokes~$I$ image 
of the PKS~1136$-$135 jet, shown with contours and polarization vectors taken from the radio $8.5\GHz$ image.  
At bottom (Figure~2b), the $8.5\GHz$  Stokes~$I$ image  image of the PKS~1136$-$135 jet,
shown with contours and polarization vectors taken from the galaxy-subtracted {\it HST} F555W Stokes~$I$ image.  
In both panels, the size of the polarization vector indicates the degree of polarization, with
a 1 arcsecond long vector representing $\Pi_{\mathrm{Radio}}=40\%$.  The direction of the vectors indicates
the direction of the inferred magnetic field (i.e., 90 degrees from the values reported in Table~2).  The 
contours are spaced by multiples of $\sqrt 2$.
See \S\S~3.1,~3.2 for discussion.}

\label{fig:opol}
\end{figure}

Virtually every jet region is extended in our deep F555W image.  Knot~$\alpha$ is revealed 
to be more than an arcsecond long.  The brightest X-ray and optical emission comes 
predominantly, but not entirely from the
upstream portion of the knot, while the radio emission comes almost entirely from its 
downstream end. For this reason, we measured the flux in knot~$\alpha$ from two regions, 
one restricted to the region seen in optical, and the second restricted to the region seen in 
radio (called $\alpha_{\mathrm{Radio}}$ in Tables~1-3).  These two regions will also be broken out in our 
discussion of the broad-band spectrum (Sections 4.2-4.3) of jet regions.  Also of note is the fact that the 
optical and X-ray emission from knots~A and~B is not co-located with the radio maxima, with
the maxima in the optical and X-ray being located upstream (i.e., closer to the quasar) than
the maxima in the radio.  These features are also much more compact in the optical than in the 
radio, and have been noted before in other jets across a wide variety of luminosity classes (e.g., PKS~1127$-$145 -- \citealt{Siemiginowska:07}, 3C~353 -- \citealt{Kataoka:08}, Cen~A -- \citealt{Hardcastle:07}, and M87 -- \citealt{Perlman:05})
%
The optically seen region of knot~C also appears to have a different
morphology in the optical than in the radio, with emission being extended along a quasi-linear
feature inclined at about 45$^\circ$ angle from the local jet direction (Figure~4).  It is unclear what this 
feature represents, but when combined with the bend that is observed at knot~D the jet in this
region appears to have a coiled appearance in the optical.  A look at the $22\GHz$ radio image
also reveals a coiled appearance in this region, but with a second region in knot~C, not seen
in the optical, that is inclined at a $\sim 90^\circ$ angle from the northern one, converging in a 
``V'' shape at the downstream end of the two regions.  The optical and X-ray images show the 
same decrease in flux with distance from the core from knot~B to knot~E; however, the optical 
flux from the hotspot is considerably brighter relative to knots~D and~E than seen in the X-rays.

\begin{figure}[h!]
  \includegraphics[height=.37\textheight,angle=270]{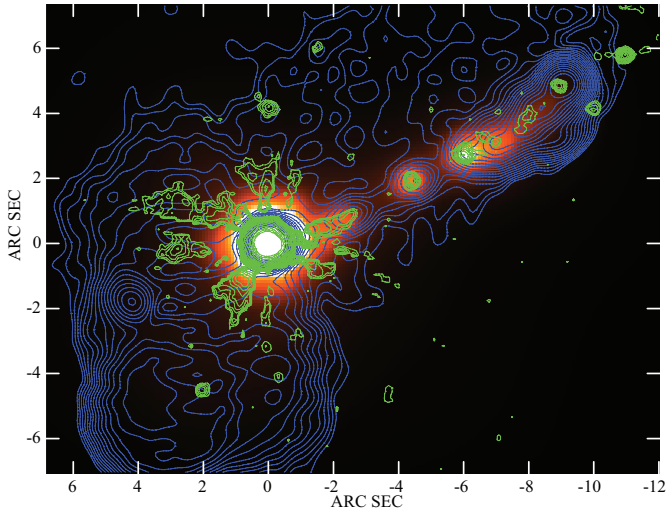}
  \caption{The {\it Chandra} X-ray image of the jet of PKS~1136$-$135 (color), with contours taken 
  from the {\it HST} F555W data (green) and radio $8.5\GHz$ data (blue).  The differences in morphology 
  between the three bands is apparent. See \S\S~3.1,~3.2 for discussion.}
\label{fig:3color}
\end{figure}

\begin{figure}[h!]
 \includegraphics[height=.37\textheight,angle=270]{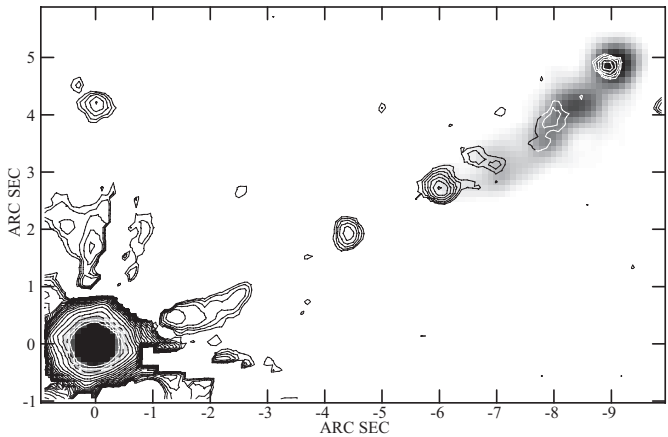}
 \caption{The $22\GHz$, VLA A-array image is shown, in greyscale, with contours taken from the 
 {\it HST} F555W image.}
\end{figure} 

\subsection{Polarimetry}

As already mentioned, Figure~2 shows the comparison of radio and optical polarimetry of the
PKS~1136$-$135 jet. 
The majority of the radio-bright knots in the jet (Figure 2b) are seen to be 
significantly polarized in the optical, with some regions having values of $\Pi$ consistent
with the theoretical maximum of $\sim 70\%$ for a perfectly ordered magnetic field.    The polarization 
vectors display definite structure, with a rotation of about 20$^\circ$ in $\chi$ seen between knot~A and 
the knot~C-D-E complex. 
 
The radio polarization map is also interesting.  The polarization in the jet interior is considerably 
lower, typically  5-10\%, than seen at the northern and southern edges of the jet, where polarizations of 
20-30\% are seen.  This trend, which is commonly seen in other quasar jets \citep[e.g.,][]{roberts:13} can be seen everywhere that the jet is resolved across its width in the $8.5\GHz$ polarization map.  Even aside from this trend, the polarization of all jet regions is lower 
in the radio than in the optical, 
with the difference being a factor of three in knot~A and a factors of 5-20 in knots~C, D and~E.  


The orientation
of the inferred magnetic field vectors (Table~2, Figure~2a) 
in the radio is generally aligned with the local jet direction, with the 
main differences being in knot~B, which displays a 90$^\circ$ flip in the magnetic field direction at its upstream
end, and in knot~C, where the vectors change direction by about 70$^\circ$ (Table~2; see also below).
In the knots where significant polarizations are detected in both bands, 
we see significant differences in the inferred magnetic field direction of 
the radio and optical emission. In knot~A, these differences may not be statistically
significant because of the fact that the northern and southern ends of the knot have the same
PA as seen in the radio.  Higher resolution radio data are needed to resolve this issue.
The differences are very significant, however, in knots~C, D and~E, as can be seen in Table~2 and also by comparing Figures~2a and~2b.  Interestingly, in knots~C and~D the inferred magnetic field
direction in the optical appears to be oriented 90$^\circ$ from the feature seen in the optical, with a PA more similar to the 
southern feature not seen in the optical.  By comparison, the radio polarization map in this region
shows near-zero polarization, perhaps due to cancellation of vectors from the two ``arms'' of this V-shaped
region, which (as can be seen from the regions north and south of the optically bright feature) have 
polarization vectors that differ in direction by 90$^\circ$.  Also of note here is knot~B, which in the radio 
map shows a radio position angle different from the dominant jet 
direction by about 60$^\circ$.  By comparison, in the optical knot~B is formally not polarized, but a 
closer look at the vectors on Figure~2 reveals that this is so only because there appear to be multiple 
magnetic field orientations in that region, leading to a non-significant results when the Stokes parameters
are added. 

Another notable feature in the radio map is the high polarization and complex characteristics of the 
southern lobe and hot spot, as well as the apparent ``sheath'' to the north of the jet.  These regions, 
which are not seen in the optical image, display very similar X-ray and radio morphologies.   The northern
``sheath'' displays a diffuse morphology.   The radio polarization vectors in this region, where seen, are oriented 
very nearly north-south, suggesting a flow out of the jet, perhaps similar to the sheath suggested for 
3C~345 and other jets \citep{roberts:13}.  By contrast, in the southern lobe the morphology in the radio is dominated by three
fairly bright features.  The first of these is a hotspot about $4.5''$
to the southeast of the quasar that is highly polarized ($\sim 40\%$) and displays a magnetic field oriented about
$10^\circ$ from north-south, features that are strikingly different from those seen in the northern hotspot.  
The second dominant feature of the southern radio structure 
are two bubble-like features that are edge
brightened and display  higher polarizations around their edges (about $30\%$), with inferred magnetic field vectors that are
correlated with the position angle of the local ``wall'' of the bubble.  Much lower polarizations are seen within 
the bubble interior as well as in the region between the hotspot and bubble, where the polarization is consistent
with zero.  

\section{Discussion}
 
The optical and X-ray emission from the PKS 1136--135 jet has been modeled with both 
synchrotron and IC/CMB models.  Early work in \citet{Sambruna:02} based on a 10 ks {\it Chandra} 
observation and a single wide-band {\it HST} image, noted that knot A had a rather 
different broadband SED than other components in the PKS 1136--135 jet.  These characteristics 
led them to favor the synchrotron model for the X-ray emission of knot A but suggest that other mechanisms prevailed in the 
other knots.  Later work \citep{Sambruna:06, Tavecchio:06} based on a much deeper {\it Chandra}
image and three  {\it HST} images, ruled out the simple, one-component synchrotron model
fit of \citet{Sambruna:02}, and favored the IC/CMB mechanism for the X-ray 
emission of all the jet components.   \citet{Uchiyama:07}, however, presented a viable alternative 
interpretation for the X-ray and optical emission as synchrotron emission from a second, high-energy 
particle population. 

In this light, the detection of high optical polarization in four regions of the PKS 1136--135 jet
is highly interesting, particularly since recent work
\citep{McNamara:09, Uchiyama:08,Uchiyama:11}
predicts that  IC/CMB emission should be unpolarized, reflecting 
the unpolarized nature of the seed photons.  This is the first clear detection of 
high optical polarization in any quasar jet region, although in the 3C~273 jet early {\it HST}
observations\citep{Thomson:93} found similarly high polarizations that were inconsistent 
with much higher signal-to-noise ground-based observations\citep{Roser:91}, and cannot 
be explained by just the difference in resolution.  Th
It has important implications for the origin of the high-energy emission in this jet,
The high polarization that we measure in the optical jet of PKS 1136-135 
has important implications for the origin of the high-energy emission, as discussed 
below.   We concentrate particularly on knot A, which is the most challenging case, although  
we will also discuss the other jet regions in depth.

\subsection{Implications of the High Optical Polarization of the Jet Regions}

IC/CMB radiation by highly relativistic electrons 
with $\gamma \gg 1$ should be unpolarized.
If instead the scattering electrons are cold ($\gamma \simeq 1$), 
the emission is in the regime of ``bulk Comptonization" and it is in principle 
highly polarized \citep{Begelman:87}.  To produce IC/CMB radiation at optical wavelengths,
the scattering electrons need to be only mildly relativistic, 
with $\gamma \sim 1\mbox{--}3$.  
Here we investigate whether polarized, bulk IC/CMB emission 
could explain the high polarizations we detect.
To calculate the intensity and polarization of the IC/CMB emission, we follow 
the prescription presented in \cite{Uchiyama:11}. The model is described by 
the electron energy distribution $n_e (\gamma )$, 
a knot radius in the jet frame of reference ($r_{\rm b}$), 
the bulk Lorentz factor of the jet ($\Gamma$), and 
the Doppler factor of the jet ($\delta$). 
We adopt $r_{\rm b} =1\ \rm kpc$, which is unimportant when we calculate polarization. 
It is a common practice to  assume  a  power law for 
the energy distribution of the electron density:
\begin{equation} 
n_e(\gamma) = \left\{
\begin{array}{cc}
k \gamma^{-s}
 &  \mbox{for}\ \gamma \geq \gamma_{\rm min}, \\
0 
 &  \mbox{otherwise}, 
\end{array}  \right.
\end{equation}
where we adopt $s = 2.4$ \citep{Uchiyama:07} and $\gamma_{\rm min} = 1.2$.   
We set $k$ by assuming that the X-ray flux is attributable to the IC/CMB emission (see \S~4.2).

\begin{figure}[ht] 
\epsscale{0.47}
\includegraphics[height=.3\textheight]{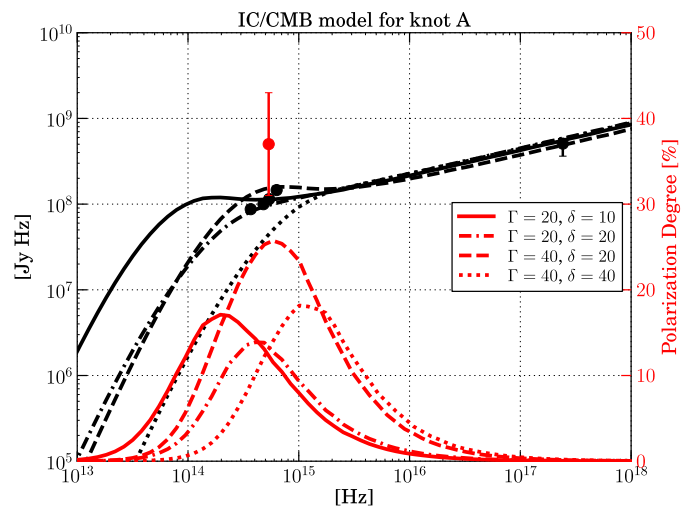}
\caption{Intensity (black curves) and polarization degree (red curves) of the IC/CMB emission from a 
relativistic jet as a function of observing frequency. 
 Data points present the \emph{HST} results for knot~A obtained in this paper.   We show plots for jet 
 bulk Lorentz factor $\Gamma=20$ and 40, with beaming factor $\delta=\Gamma$ and $\delta=\Gamma/2$.
See \S~4.1 for discussion.
 \label{fig:ICCMB}}
\end{figure}

In Figure~\ref{fig:ICCMB}, the photometric and polarimetric data for knot~A are 
compared with the IC/CMB models with $\Gamma = 20$ and 40.  We consider two cases of 
the Doppler factor,  $\delta = \Gamma$ and $\delta = \Gamma /2$, for an assumed value 
of $\Gamma$. 
Generally, the case of $\delta = \Gamma /2$ provides a higher polarization degree 
than $\delta = \Gamma $, since the scattering angle in the jet frame is optimal 
in the former case. 
The highest degree of polarization is $\Pi_{\rm max} = 26\%$ for $\delta = \Gamma /2$ and $\Gamma=40$, and
the position of the peak polarization  shifts toward higher frequencies for higher $\Gamma$. 
At the \emph{HST} band, the polarization degree ranges from $\Pi \simeq 12\%$ to $\Pi \simeq 25\%$. 
The \emph{HST} observations presented in this paper give $\Pi = 37 \pm 6 \%$.  
This can be reconciled at the $2\sigma$ level with the IC/CMB model for $\Gamma \simeq 40$, $\delta \simeq 20$ and 
$\gamma_{\rm min} \la 1.2$ , but is inconsistent with the lower $\Gamma$
and lower $\delta$ models. 
We note that the polarization vector measured with the \emph{HST} 
 is close to perpendicular to the jet axis, and therefore 
the polarization direction of the IC/CMB emission roughly coincides with that of 
the \emph{HST} measurement. 

The parameter space allowed for the IC/CMB model 
is thus tightly constrained by the optical polarization measurements, and is restricted to a very high jet 
$\Gamma\geq 40$ and 
beaming parameter $\delta \sim 20$.  The large value of $\Gamma$ is required by (a) the need to shift the lowest end of the inverse-
Compton ``hump'' into the optical (these photons would have seed electrons in the bulk, thermal tail of the EED, and 
would otherwise peak in the near-to-mid infrared, as discussed in \citealp{Georg:05}), and (b) the 
need to beam the IC/CMB sufficiently so that the bulk Compton emission dominates over the exponential synchrotron tail. 
The required beaming parameters also limit us to a viewing angle within $3^\circ$ of the jet axis.  
Finding such a combination~of jet parameters purely by chance is unlikely.  For example, one can calculate the
chance probability of finding a single jet with $\Gamma>40$ (see below) 
in MOJAVE, assuming that the Lorentz factors of the sample 
range from 1.25 to 60 with a power-law index of $k=-1.5$ \citep{Lister:09, Cara:08}, to be 3.8\%. While such a 
probability is not prohibitively low, we would expect that such a source would have properties typical of other high-$\delta$ sources, e.g., a flat radio spectrum, high integrated optical polarization and OVV "blazar" type variability, neither of which is present in PKS~1136$-$135.  




Other knots are also interesting to discuss in this regard.  Knots ~B and $\alpha$, which have 
fairly similar SEDs to 
that seen in knot~A but low ($< 15\%$) optical polarization and steeper optical spectrum, 
do not challenge existing models strongly so the polarization 
data cannot further constrain its optical-to-X-ray emission mechanism.  Both the IC/CMB and 
synchrotron models remain plausible for knots B and $\alpha$.
 Knots~C, D, and~E, by contrast, have high 
optical polarization.  Of these, two (knots~C and~D) have steep optical spectra ($\alpha_O \sim 2.5$) 
and optical-to-X-ray SEDs 
\citep{Sambruna:06} consistent with optical-to-X-ray emission originating in synchrotron radiation from 
the same population
of particles responsible for the lower-frequency (radio through IR) emission.  The third of these 
(knot~E) has $\alpha_O$ very 
similar to what is seen in the radio-through IR, suggesting that its optical emission originates in the 
same electron population
as the radio through IR emission, and additionally a much higher break frequency than seen in the 
other knots.  

\subsection{Modeling of the Spectral Energy Distribution}

Our observations place tight limitations on the parameter space available for synchrotron-inverse 
Compton emission models of knot~A.
Taking into account the requirements of high polarization, we now fit a 
synchrotron-inverse-Compton model to the multi-wavelength data for knot~A (Table~1).   We follow the prescription used in 
\citet{Perlman:11} for setting values of jet $\Gamma$, $\delta$ and viewing angle $\theta$ and attempt to match the 
observed broad-band emission.  The hard optical spectrum of knot~A is quite constraining in this regard.  For example, using
the low end of the range of $\Gamma$ values allowed by the polarization modeling done in \S~4.1 results in a flat 
optical spectrum, and also does not allow us to pick up the peak of the polarized emission, both necessary given these 
observations.  
We therefore found it necessary to direct the modeling towards higher values of $\Gamma$ and use $\delta=
\Gamma/2$ to maximize the polarized flux. This further restricts the region of parameter space open to an IC/CMB model. 

As shown in Figure~\ref{SEDs}, it is possible to fit the broadband SED of the X-ray bright 
knot~A   with both a synchrotron and an IC/CMB  model.  The IC/CMB fit (top) requires a bulk Lorentz factor 
$\Gamma=40$ and a Doppler factor $\delta=20$ (i.e., $\theta=2.48^\circ$). This is a much more 
extreme set of parameters than previously required for the PKS 1136--135 jet in the IC/CMB model \citep{Sambruna:06}.

A simple power law electron energy distribution (EED) cannot accommodate the data. The reason for this is that a single power law EED  that fits the radio would produce an IC/CMB spectrum that would be very hard and, if fitted the X-ray flux, would extend from there below the  observed optical fluxes. For this reason we adopted a broken power law that is harder {\sl above} a break energy: 
$n(\gamma)\propto \gamma^{-2.7}$ for $\gamma_{min}\leq \gamma \leq \gamma_{break}$   and 
$n(\gamma)\propto \gamma^{-2.3}$ for $\gamma_{break} <\gamma \leq \gamma_{max}$, with 
$\gamma_{min}=1.6$,  $\gamma_{break}=160$, and $\gamma_{max}=3 \times 10^5$. Such a break, with the EED hardening above a given energy, although  not usually discussed in the literature, is generally expected at the energy where the high energy power law component of the EED  starts to dominate over the  low energy relativistic Maxwellian component, as relativistic particle-in-cell simulations show \citep{spitkovsky08}.  It also necessitates a slightly higher $\gamma_{min}$, as the 
lower value of \S~4.1 would overproduce the optical-UV tail of the high-energy component.
The jet kinetic power, assuming one cold proton per lepton, is $\sim 34 $ times the Eddington luminosity of  a $10^9$ M$_\odot$ black hole.
For a projected jet length of $\approx 11''$  
and jet orientation $\theta = 2.48^\circ$ (see above), the deprojected length of the jet is 1.63 Mpc, comparable with the largest known jets \citep{konar04}.
  
In the bottom panel of Figure~\ref{SEDs} we plot a two-synchrotron component SED for knot~A. For this representation we chose $\Gamma=\delta=2$, corresponding to a jet angle to the line of sight of $\theta=30^\circ$. The first population of electrons, reproducing the radio to optical SED is a power law with $\gamma_{min,1}=100$, $\gamma_{max,1}=
7 \times 10^5$, electron index $2.4$ and power $L_{e,1}=1.2 \times 10^{45}$ erg s$^{-1}$. The second component 
reproducing the optical to X-ray SED, which could form as a result of a continuous acceleration piling up radiating 
ultra relativistic electrons around the maximum energies available in the acceleration process \citep[e.g.,][]{stawarz08} 
is a power law with $\gamma_{min,2}=3\times 10^6$, $\gamma_{max,2}=
2 \times 10^8$, electron index $2.0$ and power $L_{e,2}=2.4 \times 10^{43}$ erg s$^{-1}$.

The magnetic field set to the equipartition value is  $B=4.7 \times 10^{-5}$ G. The jet kinetic power, assuming one cold proton per lepton, is
$L_{jet}=8.8 \times 10^{45}$ erg s$^{-1}$,  $\sim 7\% $ of the Eddington luminosity of  a $10^9$ M$_\odot$ black hole. The deprojected  length of the jet is $140$ kpc.

 This modeling of the SEDs shows that 
the IC/CMB model faces significant issues, as one needs
very high beaming factors and small jet viewing angle to push IC/CMB from the $\gamma \sim 1$ 
electrons to high enough frequencies to explain the high polarization observed in the optical. 
However, given that PKS~1136$-$135 does 
not exhibit properties typical
of blazars, such as rapid variability, flat radio spectrum and high core 
polarization (\S~1), these parameters are  not favored. 
Thus, while the IC/CMB model for the X-ray emission is restricted to a small 
and unlikely range of parameter space, it is not completely ruled out.

\subsection{An additional diagnostic: the slope of the optical-UV spectrum}

\begin{figure}[ht!]
\begin{center}
\includegraphics[width=0.49\textwidth]{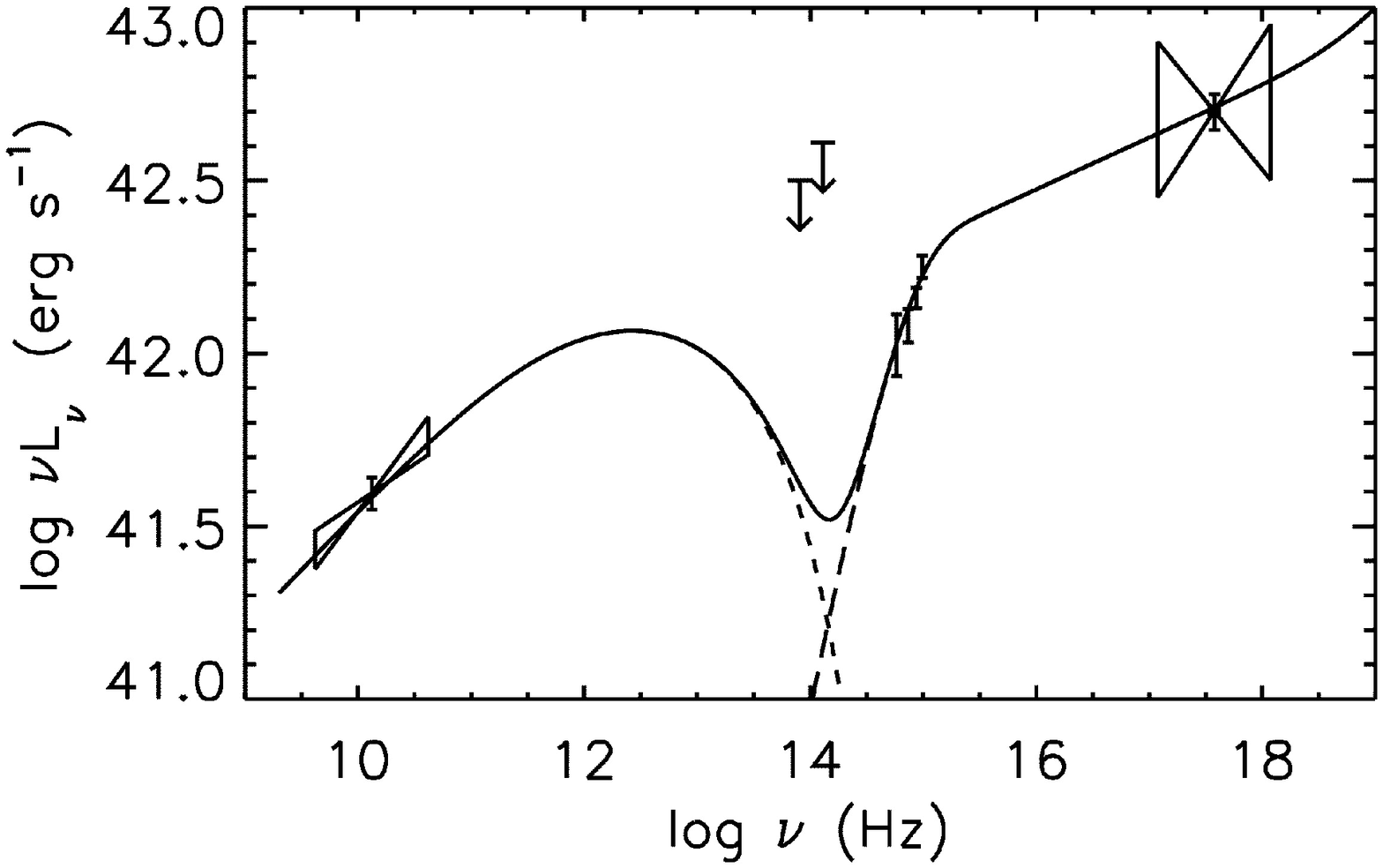}
\includegraphics[width=0.49\textwidth]{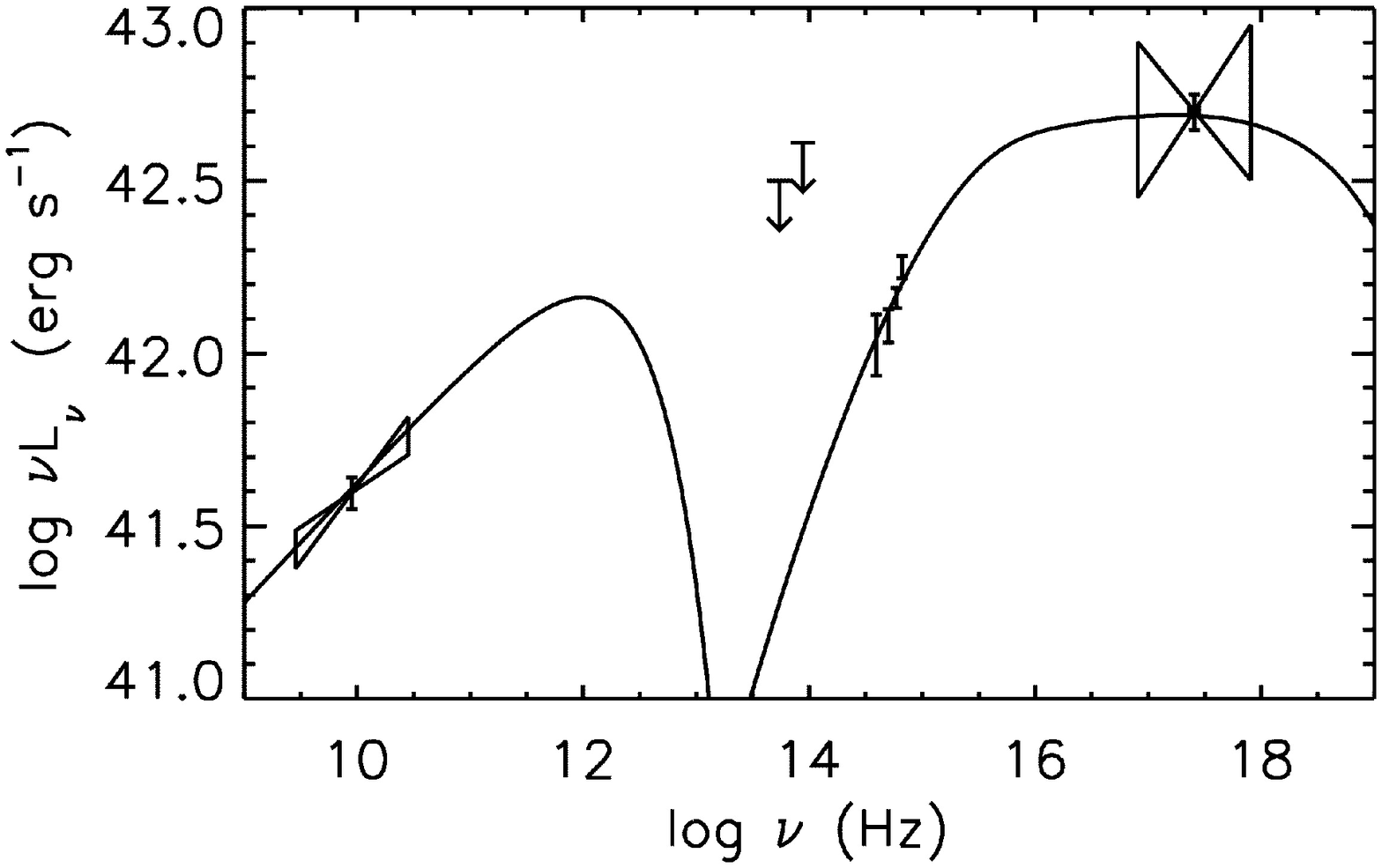}
\end{center}
\caption{ Fits of the  knot~A data of PKS~1136$-$135 with the IC/CMB (top) and two-synchrotron component (bottom) models. Both fits are 
reasonably good. They correspond, however, to extremely different jets  (see text).}
\label{SEDs}
\end{figure}

Further progress can be made to narrow down the constraints on all the emission models by follow-up 
{\it HST} observations, which would allow further study of their optical-UV spectral shapes.  In knot~A, 
this could strengthen further the case against IC/CMB.
 This is based on   the different  slopes  the inverse Compton and synchrotron  low energy spectral 
 tails have.  The low energy tail of the X-ray component, produced by the lowest energy electrons of 
 the EED,   will have a spectral index  $\alpha=-1/3$ for the synchrotron and $\alpha=-1$ for the IC/
 CMB mechanism  \citep[e.g.,][]{stawarz08,dermer09}.
 If this tail extends to the  optical - UV part of the spectrum, a measurement of the spectral index can be used to identify the 
 emission mechanism.
 In the case of PKS~1136$-$135  the optical-UV spectrum of knot~A is indeed hard and is  part of the low energy tail of the X-ray component (see Figure~\ref{SEDs}).

\begin{figure}[h!]
\begin{center}
\includegraphics[width=0.5\textwidth]{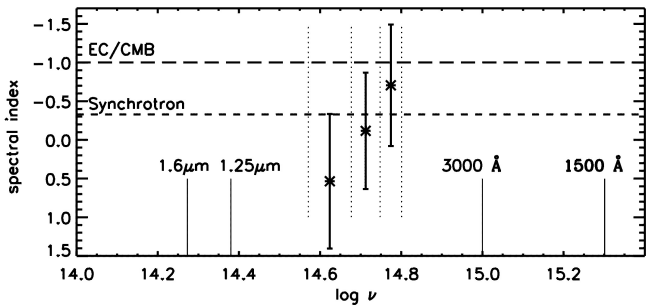}
\end{center}
\caption{The three two-point spectral  indexes of knot~A of PKS~1136$-$135. The long and short dash horizontal lines represent the spectral index expected from the low energy tail of  IC/CMB and synchrotron emission respectively. The thin vertical dotted lines mark the four wavelengths for which we have data. The four wavelengths at which  future {\emph HST} observations could extend the spectrum are also shown.}
\label{fig:alphas}
\end{figure}

  The existing fluxes (from F814W,  F625W, F555W and F475W) cover a factor of less than~2 in 
  frequency. 
  We plot in Figure~\ref{fig:alphas} the three two-point spectral indexes we form from these  fluxes. 
  As can be seen, the spectrum appears to
harden with frequency. This suggests that at low frequencies the observed flux is the sum of the hard  low energy tail of the X-ray component and the soft high energy tail of the radio - IR - optical synchrotron component. As we move to higher frequencies the contribution of this soft tail decreases and the spectral index we observe is closer to the actual spectral index of the low energy tail of the X-ray component. To exclude an emission mechanism we need the observed spectral index at any part of the optical-UV-spectrum to be harder than $-1/3$ for synchrotron and $-1$ for IC/CMB.
Our existing data do not permit this, although the hardening of the spectrum at higher frequencies leaves open the possibility that at  UV energies it may be possible to exclude the synchrotron mechanism.
This would force us to accept IC/CMB with all of its important consequences of fast and powerful jets.
On the other hand, the spectral index from the three highest energies (F625W, F555W and F475W) is $0.37\pm0.20$, which is in  agreement with the synchrotron case.   Additional  {\emph HST} observations at $1.6 \,\mu$m, $1.25\, \mu$m, 3000 \AA~ and 1500 \AA, would be extremely useful
in characterizing the IR-optical-UV spectrum and constraining further the optical-UV-X-ray emission mechanism.

Future observations can also narrow down the constraints for the other knots.  The spectrum of knot~$\alpha$ 
is also consistent with either synchrotron emission from a second, high-energy electron population, but the constraints on 
IC/CMB radiation are much less severe because of its low optical polarization.  Additional {\emph HST} observations could allow
us to detect a spectral break that could decide between these two interpretations. For knot~B, the current data are consistent 
with either a significantly higher $\nu_{\mathrm{break}}$ in the synchrotron emission or the optical emission could be from IC/CMB 
radiation.  Both of these possibilities would be constrained by observations in the near-IR and UV, as each of these two 
models would predict spectral breaks that would be detectable either in the UV or near-IR.  Finally, for knot~E, if the optical 
emission is from the same electron population responsible for the radio-IR component, the observed X-ray flux and spectral index (Tables~2,~3) predict a steepening in its spectral index towards the UV.  
%

\vspace{0.1cm}


\subsection{The Jet Power}

\begin{figure}
\includegraphics[height=.28\textheight]{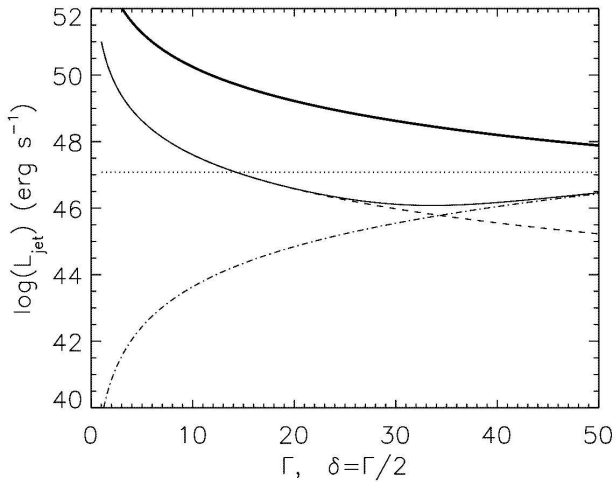}
\caption{Power requirements for knot~A, using the observed spectral energy distribution and the IC/CMB interpretation. 
Tracks are shown for the electron power (dot dashed
line), the Poynting flux (broken line), the total power for leptons
only (thin solid line) and the total power for one  proton per
electron (thick solid line).  Beaming factor $\delta$ is constrained to be equal to $\Gamma/2$,
but for a given $\Gamma$, the magnetic field required is then set by the ratio of radio to 
X-ray flux \citep[see e.g.,][and references therein]{Perlman:11}.  See \S~4.2 for discussion.}
\label{fig:jetpower}
\end{figure}

Another constraint on the nature of the observed optical and X-ray emission and also the nature of the jet can be achieved
by looking at the jet power implied by such a model.  In Figure~{\ref{fig:jetpower}} we show the result of modeling the power 
requirements for knot~A, showing tracks for electron power, Poynting flux, leptons only and total for one proton per electron.
To produce this plot we followed the prescription of \citet{Mehta:09} and \citet{Georg:05}.  In this plot we 
have required that the X-ray and radio power agree with the observations.  As can be seen, for 20 $< \Gamma <$ 50 the
leptonic power required is sub-Eddington.  In particular, for the $\Gamma=40$, $\delta=20$ model discussed above, the 
leptonic power required is 3.1 $\times 10^{46}$ erg/s, about 0.3 $\times$ the Eddington luminosity of PKS~1136$-$135's 
black hole.  However, if we require one cold proton 
per lepton, the power requirement is much more extreme, about 10$\times$ Eddington. As discussed above, we do not 
favor this model for a variety of reasons, but if we go to a different region of parameter space which appears more 
reasonable given the observed properties of PKS~1136$-$135 (i.e., lower values of $\Gamma$ and $\delta$), the total power
requirement goes up significantly and is in excess of Eddington even for a lepton only jet at $\Gamma<10$.  Furthermore, 
in such a case we could not reproduce the observed optical polarization as in that case the optical emission 
would have to come from IC/CMB with particles at higher Lorentz factors $\gamma$.  

The alternative to the IC/CMB model is to 
produce the optical and X-ray emissions in a  second, high-energy 
synchrotron component.  This model is 
attractive because one is not at all restricted in the choice of bulk Lorentz factors $\Gamma$ or 
viewing angle $\theta$ since there is no requirement to produce the highly polarized optical 
emission with IC/CMB off the very lowest energy, $\gamma \sim 1$ particles. Such a model would 
easily explain the high polarizations we observe and also have the advantage of lowering the jet 
power requirement by approximately $\sqrt{m_p/m_e}$, 
as here the 
ratio of jet radio to X-ray luminosity would not be fixed for a given set of parameters. 

\section{Conclusions}

The origin of the high-energy emission in quasar jets is a long-standing issue, dating back to the discovery of 
X-ray emission from the jet of 3C~273 \citep{Willingale:81}.   Early work \citep{Harris:87} found that it was 
difficult to interpret the X-ray emission as either an extension of the radio-optical emission or due to the 
synchrotron self-Compton process.  Later work, following the launch of the {\it Chandra} X-ray 
Observatory, centered around IC/CMB emission \citep[e.g.,][]{Schwartz:00, Celotti:01, Marshall:01}.  
This process, like SSC, is attractive because it is mandatory --
the mere presence of high-energy particles ensures emission from IC/CMB at some level.  However, 
our ability to observe this component, combined with the strength of the component itself, is highly 
geometry
dependent.  In particular, for the IC/CMB component to dominate the observed X-ray emission, the jet must be 
strongly beamed, and remain relativistic for tens to hundreds of kiloparsecs from the active nucleus.  In some 
objects, such as PKS~0637$-$752, where strong gamma-ray emission and/or other blazar properties are observed, as
well as a lack of a continuous optical-to-X-ray component \citep[see e.g.,][]{Mehta:09}, this
seems to be the leading hypothesis.  However, in several objects, such as 3C~273 and PKS~1136$-$135, the lack of 
evidence for strong beaming and the presence of optical to X-ray emission from a single spectral component, has 
reinvigorated the possibility that the X-ray and optical emission may be
synchrotron emission from a second, high-energy population of electrons \citep{Jester:06, Jester:07, Uchiyama:06}.

The observations discussed in this paper argue strongly against the IC/CMB process as being 
dominant in the X-ray band, at least for the 
case of the jet of PKS~1136$-$135.  While the SED of knot A shows that the X-ray and optical 
emission are clearly linked, the high polarization we observe in knot~A and other components 
require a very highly beamed jet, with bulk Lorentz factor $\Gamma \geq 30$ and viewing 
angle within 3$^\circ$ of the jet axis.  These constraints (much tighter than previous work; 
\citealt{Sambruna:06, Tavecchio:06}) are a result of the need for the polarized optical
emission to come from the very lowest energy ($\gamma \sim 1$) electrons, as IC/CMB emission 
from higher energy particles would be unpolarized.   They are also substantially tighter than 
typically required for the IC/CMB process, as where 
polarization data are not present there is no requirement for the observed optical emission (if any) to 
come from bulk Compton emission from $\gamma \sim 1$ electrons (as in the case of PKS~0637$-
$752, \citealt{Mehta:09}). 
The required configuration is highly unlikely given the observed properties of PKS~1136$-$135, which 
has a steep radio spectrum and displays neither  rapid 
variability or high integrated optical polarization, unlike blazars, the more typical, high-$\delta$ source.  
Furthermore, the energetic demands of such a jet are extreme:  
if we require one proton per radiating lepton, the jet power must be at least 10 $\times$ the Eddington 
luminosity of PKS~1136$-$135's black hole, and such a configuration might result in Faraday 
depolarization in the radio \citep{jones77}.

The observations presented here instead favor a more complicated story.  While the IC/CMB 
process is mandatory, synchrotron emission from a second, high-energy particle population 
is the favored interpretation for knot A's X-ray emission, which is spectrally linked to the polarized 
optical emissions.  Synchrotron emission is also required to explain the highly polarized optical
emissions of knots C, D and E.   Two of those knots (C and D) show optical spectra that are 
decreasing in $\nu F_\nu$ and do not appear spectrally linked to the X-ray emission.  For those knots, 
we cannot use the optical polarization characteristics to infer conclusions about the nature of the 
X-ray emission.  Knot E, however, has an SED much more similar to knot A and is also highly
polarized.  A similar explanation for its X-ray emission seems likely.  The two-component synchrotron
model, first suggested for the jet of PKS 1136--135 by \citet{Uchiyama:07}, requires the 
high-energy particle population to be spectrally distinct from that seen at lower (radio-infrared) 
energies, but does not specify whether it is spatially co-located with the lower-energy one.  
Discriminating between these two possibilities requires further, high-angular resolution work, and
while we do see significant differences between the X-ray, optical and radio morphologies 
(e.g., Figure 3), the resolution in the X-rays is insufficient to comment further.

\clearpage

\begin{acknowledgments}

The National Radio Astronomy Observatory is a facility of the National Science
 Foundation operated under cooperative agreement by Associated Universities, Inc.
E. S. P., M. C., and M. G. acknowledge support from NASA grants NNG05-GD63DG at UMBC and NNX07-AM17G at FIT, 
and {\emph HST} grant STGO-11138.
C. C. C. was supported at NRL by a Karles' Fellowship and NASA DPR S-15633-Y.  L. S. was supported by Polish NSC 
grant DEC-2012/04/A/ST9/00083.

\end{acknowledgments}


%
%
%
%
%
%



\end{document}

%% file: table1.tex
\begin{center}
\begin{deluxetable*}{cccccccccc}
\tabletypesize{\scriptsize}
\tablewidth{0pt}
\tablecolumns{10}
\tablecaption{Flux Densities ($F_\nu$) of Jet Features}
\tablehead{
\colhead{Feature} & \colhead{8.5 GHz$^a$} & \colhead{22 GHz$^a$} & \colhead{5.8 $\mu$m$^b$} & \colhead{3.6 $\mu$m$^b$} &  \colhead{F814W} & \colhead{F625W} & \colhead{F555W} & \colhead{F475W}  & \colhead{1 keV$^a$}  \\
		  & \colhead{(mJy)} & \colhead{(mJy)} &\colhead{(mJy)} &\colhead{(mJy)} &\colhead{(nJy)} & \colhead{(nJy)} & \colhead{(nJy)} & \colhead{(nJy)} & \colhead{(nJy)} }
\startdata
$\alpha$         & $3.0  \pm 0.3$ &  $ 1.5  \pm 0.6 $	&		            &    			      & $337 \pm 59$  & $ 330 \pm 50$	& $ 358 \pm 16 $ & $ 362 \pm 42 $  & $ 1.9 \pm 0.2 $ \\
$\alpha_{\mathrm{Radio}}$ & $3.0  \pm 0.3$	&  $ 1.5  \pm 0.6 $	&		            & 				      & $320 \pm 60$  & $ 334 \pm 50$	& $ 261 \pm 15 $ & $ 249 \pm 42 $  & $ 1.9 \pm 0.2 $ \\
A	               & $3.8  \pm 0.4$	&  $ 2.0  \pm 0.6 $ & $<5$          & $< 4$ 			  & $237 \pm 48 $ & $ 208 \pm 23$ & $ 212 \pm 14 $ & $ 231 \pm 17 $  & $ 1.7 \pm 0.2 $ \\
B	               & $9.3  \pm 0.9$	&  $ 3.2  \pm 0.6 $ & $4.8 \pm 2.2$ & $3.8 \pm 1.9$ & $535 \pm 71$  & $ 396 \pm 35$ & $ 365 \pm 24 $ & $ 320 \pm 42 $  & $ 3.5 \pm 0.2 $ \\
C	               & $20.6 \pm 2.1$	&  $ 9.2  \pm 1.8 $ &		            & 				      & $342 \pm 91$  & $ <90^a$ 		  & $ 148 \pm 6 $  & $ <70^a $       & $ 1.8 \pm 0.2 $ \\
D	               & $29.5 \pm 3.0$	&  $ 11.9 \pm 2.4 $ &	 	            &				        & $390 \pm 130$ & $ 191 \pm 55$ & $ 140 \pm 17 $ & $ <80^a $  	   & $ 1.0 \pm 0.2 $ \\
E	               & $66.1 \pm 6.6$	&  $ 26.4 \pm 5.3 $ &		            & 				      & $161 \pm 52 $ & $<50^a$ 		  & $ 130 \pm 18 $ & $ <80^a $	     & $ 0.7 \pm 0.2 $ \\
D+E	             & 							  &			              &		            &	 			        & $500 \pm 150$ &				        & $ 240 \pm 30 $ &    		    	   &                 \\
C+D+E            & $116  \pm 8^{b,c}$ &  $47.5 \pm 6.1^b$ & $20 \pm 4$ 		& $9.5 \pm 2.0$ & $840 \pm 240$ & 		 	        & $ 383 \pm 31 $ &                 &                 \\
HS               & $119  \pm 12 $ &  $ 43.9 \pm 8.8 $ & $7.4 \pm 2.7$ & $4.6 \pm 2.0$ & $287 \pm 16$  & $ 159 \pm 33$ & $ 159 \pm 24$  & $ 115 \pm 26 $  &  $ < 0.6	$      \\
\enddata
\tablenotetext{a}{See Sambruna et al. (2006)}
\tablenotetext{b}{From Uchiyama et al. (2007)}
\tablenotetext{c}{Errors recomputed in quadrature}
\end{deluxetable*}
\end{center}

%% file: table2.tex
\begin{center}
\begin{deluxetable}{cccccc}
\tablecolumns{6}
\tablewidth{0pt}
\tablecaption{Optical and Radio Polarimetry of Jet Features}
\tablehead{
\colhead{Feature} & \colhead{R ($''$)} & \colhead{$\Pi_{O}$, \%} & \colhead{$\mathrm{\Xi}_{O}^a$} & \colhead{$\Pi_{R}$, \%} & \colhead{$\mathrm{\Xi}_{R}^a$}
}
\startdata
$\alpha$ &	2.7	& $< 15^b$ 	& ....	     & $21 \pm 2$ & $ 17 \pm 5$ \\
$\alpha_{\mathrm{Radio}}$ & 	& $< 15^b$	& ....       & $22 \pm 2$ & $ 19 \pm 5$ \\
A	&	4.6	& $37 \pm 6$	& $41 \pm 4$ & $11 \pm 2$ & $ 15 \pm 5$ \\
B	& 	6.5	& $< 14^b$	& ....       & $10 \pm 2$ & $ 29 \pm 5$ \\
C	& 	7.7	& $92 \pm 14$	& $65 \pm 4$ & $8 \pm 2 $ & $ -43 \pm 5$ \\
D	& 	8.6	& $53 \pm 14$   & $68 \pm 7$ & $5 \pm 2 $ & $ -31 \pm 5$ \\
E	& 	9.3	& $63 \pm 14$ 	& $68 \pm 6$ & $12 \pm 2$ & $ 15 \pm 5 $ \\
D+E	& 		& $58 \pm 11$   & $68 \pm 5$ & $8 \pm 2 $ & $ 23 \pm 5 $ \\
C+D+E   & 		& $70 \pm 9$	& $67 \pm 3$ & $8 \pm 2 $ & $ 26 \pm 5 $ \\
HS	& 	10.3	& $< 13^b$	& ....       & $9 \pm 2 $ & $ 37 \pm 5 $ \\
\enddata
\tablenotetext{a}{Electric field vector position angle.}
\tablenotetext{b}{$2 \sigma$ upper limit.}
\end{deluxetable}
\end{center}

%% file: table3.tex
\begin{center}
\begin{deluxetable}{cccc}
\tabletypesize{\scriptsize}
\tablewidth{0pt}
\tablecolumns{4}

\tablecaption{Spectral Indices ($F_\nu \propto \nu^{-\alpha}$) of Jet Features}
\tablehead{\colhead{Feature} & \colhead{$\alpha_R$$^a$} & \colhead{$\alpha_O$} & \colhead{$\alpha_X^a$} \\ }
\startdata
$\alpha$			&$0.75 \pm 0.10$ & $-0.16 \pm 0.09 $ 	& $0.9 \pm 0.4$ \\
$\alpha_{\mathrm{Radio}}$   &$0.75 \pm 0.10$ & $0.5 \pm 0.3 $		&\\
A				&$0.67 \pm 0.11$ & $0.1 \pm 0.2$	  &$1.1^{+0.3}_{-0.6}$ \\
B				&$0.81 \pm 0.13$ & $1.0 \pm 0.1$		&$1.1^{+0.2}_{-0.3}$ \\
C				&$0.66 \pm 0.09$ & $2.6 \pm 0.9$		&$0.5^{+0.3}_{-0.2} $\\
D				&$0.71 \pm 0.08$ & $2.3 \pm 0.9$		&$0.5 \pm 0.5$ \\
E				&$0.82 \pm 0.09$ & $0.5 \pm 0.90$		&$1.3^{+0.6}_{-0.5} $\\
HS		  &$0.85 \pm 0.08$ & $1.6 \pm 0.3$		&$0.7^{+0.9}_{-0.7}$ \\
\enddata
\tablenotetext{a}{See Sambruna et al. (2006)}
\end{deluxetable}
\end{center}